\documentclass[10pt,a4paper,twoside]{article}
\usepackage{epsfig}
\usepackage{baltlat6}
\usepackage{array}
\usepackage{here}
\usepackage{epstopdf}
\begin{document}
\ \ \vspace{0.5mm} \setcounter{page}{1} \vspace{8mm}


\titleb{ON THE POSSIBILITY OF THE LONG LIFETIME \\OF MOLECULAR CLOUDS}

\begin{authorl}
\authorb{A. Kasparova}{1} and
\authorb{A. Zasov}{1}
\end{authorl}

\begin{addressl}
\addressb{1}{Sternberg Astronomical Institute of Moscow State University,
\\ Universitetskii pr. 13, Moscow, 119992, Russia;
anastasya.kasparova@gmail.com, zasov@sai.msu.ru}
\end{addressl}


\begin{summary}
Arguments are given that at least a fraction of molecular clouds may live much longer time that it is usually assumed, without the transition into diffuse atomic gas (HI). We propose that star formation in these clouds may be strongly delayed and weakened by the magnetic field.
\end{summary}

\resthead{On the possibility of the long lifetime of molecular clouds} {A. Kasparova, A. Zasov}
\sectionb{1}{Introduction}
Most stars in galaxies are thought to form in dense clumps of molecular clouds (MCs), but the conditions in which MCs form and collapse are poorly understood. They may be a result either of slow gravitational contraction of gas or of collision of gaseous flows due to supersonic turbulence. Most or at least a significant part of the observed MCs are virialized or close to virial state (Heyer et al., 2009) (see however Dobbs et al. (2011) who claim that at least 50\% of them are unbound). 

At any case MCs are formed from the atomic gas and their matter becomes atomic again after their disruption and molecular dissociation. It agrees pretty well with the observed relationship between the local ratios of molecular over atomic gas mass per unit area $\Sigma_{H2}/\Sigma_{HI}$ and the gas pressure or a total surface density of gas (Blitz \& Rosolowsky, 2006, Krumholz et al., 2009). It gives evidence that the molecular and atomic gas phases are in the approximate equilibrium, although a theory says nothing about a characteristic time needed to reach it. 

It is generally accepted that even gravitationally bound MCs are short--lived objects, being easily disrupted by stellar feedback soon after the beginning of star formation due to radiation pressure, stellar winds, SN explosions and the expansion of HII regions (the analysis of different stellar feedback mechanisms see for example in Hopkins et al., 2012). Some observational arguments seem to support the short age of MCs. First, MCs are usually concentrated in spiral arms or HI filaments, where they can't stay longer than $\sim 10^7$ years. In addition, stellar clusters connected with MCs are always young (not older than several millions years old). Note however that T Tau stars of much larger ages were found in the vicinity of some currently active MCs being drifted from them (Feigelson, 1996). Murray et al. (2010) on the statistical ground showed that the lifetime of giant molecular clouds in our Galaxy is $17\pm4$ Myr or about two free-fall times. However, they considered only the most massive clouds (about $10^6 M_{\odot}$ in the mean) \textit{possessing notably} efficient star formation. It is worth noting that newborn stars may decouple from a parent cloud especially if a cloud loses a fraction of its mass or is magnetically supported.

It is natural to expect that MCs may be totally disrupted and dispersed in the case of active and efficient birth of massive stars in their interiors. The question is how common is this scenario.
\begin{center}
\begin{figure} [h!]
\includegraphics[width=9.0cm]{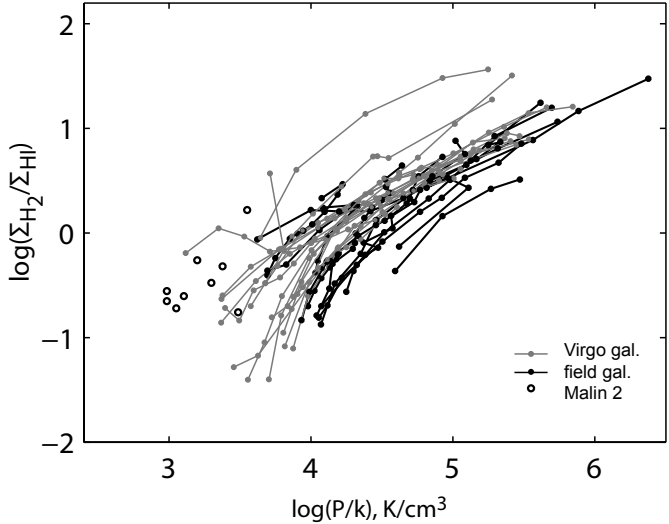}
\caption{Molecular hydrogen fraction versus turbulent gas pressure in the disk midplane for the field (black lines) and cluster (gray lines) galaxies. Malin 2 is marked by open circles.}
\end{figure}
\end{center}

\sectionb{2}{The arguments for the long life of molecular clouds}

There are several observational evidences which allow to admit that MCs are not too fragile, as it may seem, and at least a fraction of them may exist much longer than it is usually assumed without transforming molecular to HI gas. We discuss them below.

\begin{enumerate}
\item High resolution observations of CO--emission in the gas--rich galaxy M51 clearly showed that the most massive molecular clouds are not fully dissociated into atomic gas after experiencing the star formation in the spiral arms of this galaxy. Giant MCs are rather fragmented into smaller clouds upon leaving the spiral arms, and the majority of the gas remains molecular even in the inter--arm regions, so that MCs of relatively low masses reveal a weak concentration to the arms (Koda et al., 2009). Smaller MCs may collide and merge in spiral arms forming giant molecular assosiations (Egusa et al., 2010). 

\item A significant, if not overwhelming part of the observed MCs or their cores reveals the presence of the current star--forming activity. In other words, the starless MCs are rare observed (Ballesteros-Paredes \& Hartmann, 2007, Jijina et al., 1999). For example, analysis of giant MCs in M33 demonstrated that only one sixth of them had no (massive) star formation. It means that the average time required for star formation to start should be about one sixth of the total time of cloud existence (Gratier et al., 2012). Hence, either the time of formation of MCs should be extremely short, which is hard to reconcile with the models of virialized clouds, or star formation is a long-lasting process, proceeding without disrupting MCs. The other possibility which we don't consider here, is that the amount of MCs without young stars is strongly underestimated due to observational selection caused by their low CO--luminosity. The last two possibilities mean the underestimation of the generally assumed lifetime of MCs.
 
\item The most of HI--deficient spiral galaxies in clusters retain a normal content of molecular gas (see, for example, Corbelli et al., 2012), in spite of the pressure of diffuse gas being too low for the observed $\Sigma_{H2}/\Sigma_{HI}$ ratio. As an illustration, Fig. 1 shows a diagram ``Molecular gas fraction over gas pressure P/k'' calculated for the field (black lines) and Virgo spiral galaxies (gray lines). The diagram was obtained using the available data for radial gas distributions and kinematic properties of galactic disks (see the details in Kasparova, 2012). Radial profiles of the midplane pressure were calculated for gas disks with a given turbulent velocity dispersion which are in the local hydrostatic equilibrium in general gravitational fields created by gas components, stellar disk and dark halo). Open circles in the diagram mark the position of some regions in the rarefied extended disk of giant low surface brightness spiral galaxy Malin 2, based on the CO--observations by Das et al. (2010). A hydrostatic gas pressure there is extremely low due to the low surface density of very extended stellar--gaseous disk of this galaxy. As it follows from the diagram, the molecular gas fraction is unexpectedly high both in the considered Virgo galaxies and in Malin 2. Note that the star formation efficiency ($SFR/\Sigma_{H_2}$) is nearly the same, at least within the $0.5R_{25}$, for most of these Virgo galaxies (Vollmer et al., 2012). It creates a problem to explain how \textit{molecular gas} can be formed and why it was not dissociated into HI. May the molecular gas be preserved, being left from the epoch when the galaxies were gas--rich?
 
\item The another hint to the long lifetime of MCs is star formation in the intergalactic space. It is well known that star formation may occur in the local regions of tidal debris of interacting galaxies~--- close or far from their main bodies. These birthplaces are usually small and optically faint, revealing themselves by emission spectral lines and/or by UV radiation (see, for example, Neff et al., 2005, Mullan et al. 2011, Boquien et al., 2009, Karachentsev et al., 2011, de Mello et al., 2008, and references therein). Most of the young stars in the tidal tails formed in single bursts rather than resulting from continuous star formation (Neff et al., 2005). They are usually coincident with HI density enhancements in the tidal structures. The surface density of HI clouds related to such young systems may be as low as $10^{20}\ cm^{-2}$ (de Mello et al., 2008) which excludes gravitational instability. It is essential that, according to Boquien et al. (2010), the spectral energy distribution of star--forming regions in collision debris resembles more that of dusty star--forming regions in galactic disks than that of typical star-forming dwarf galaxies. One may propose that the seeds of the currently starforming clouds in these systems were expelled by tidal forces from parent galaxies parallel with HI gas. A typical time delay for the beginning of star formation for MCs which were formed in the main disk and left a galaxy, should correspond to $10^8-10^9$ years.
\end{enumerate}

The arguments given above may not be considered as the proofs of a long life of MCs. They just show that such possibility may nicely fit with observational facts. Note that the idea of the existence of a ``hidden'' cold molecular gas  which can be visible only after its heating, was claimed earlier by Pfenniger and Combes (Pfenniger \& Combes, 1994). A dynamical friction of long lived giant molecular clouds was also proposed by Surdin (1981) to account for a re-distribution of a disk matter in the central part of the Galaxy.

\sectionb{3}{On the conditions of survival of clouds}

Molecular cloud may live long and avoid a global collapse if it is magnetically supported or at least it has a very low formation rate of massive stars which is not enough to disrupt a cloud, but is high enough to support the internal turbulent motion of gas. According to Williams \& McKee (1997), for low massive MCs $(\le10^4M_\odot)$ a mean lifetime may reach $5\cdot 10^8$ yr or more even in the case of continuing star formation~--- just because the massive stars, which are the principal agents of cloud destruction, are rare. Although numerically most clouds are of low mass, most of the molecular gas is contained within the most massive clouds. For massive clouds magnetic field may provide the mechanism which significantly increases the duration of formation of gravitationally unstable cores and may reduce or forbid star formation. As numerical hydrodynamic simulations show, if a cloud is subcritical (that is relationship between magnetic and gravitational energy exceeds unit), it oscillates without forming stars till this ratio does not change (Vazques-Semadeni et al., 2011). As it was demonstrated by Heitsch et al. (2001), even the presence of the supersonic turbulence cannot help local magnetostatically supported regions to collapse. Note however, that 3D magnetohydrodynamic simulations show that a strong supersonic turbulence is needed to speed up the ambipolar diffusion (see the discussion in Kudoh \& Basu (2011).

Observations show that the densities of magnetic, turbulent and gravitational energies of clouds are usually comparable, although the measurements of magnetic field remain rather crude. Nevertheless, there are some direct evidences of the prevailing of magnetic energy. This is a discovery of non-random geometry of magnetic field inside of clouds found in M33 and in our Galaxy from the polarization of CO emission (Goldreich-Kylafis effect), which gives evidence that the B-field energy dominates turbulent energy (see Li \& Henning, 2011, and references therein). In some clouds a correlation was found between the magnetic field direction on core scales and on the scale larger than 200 pc (Li et al., 2009). All these facts give evidence that the magnetic field provides a dominant force in MCs, and that formation of dense cores may have been driven by ambipolar diffusion in strongly subcritical clouds. If the ambipolar diffusion is not efficient, subcritical molecular clouds may avoid collapse during the arbitrary long time (Bash \& Dapp, 2010; Bertram et al., 2012). It is worth nothing that not only formation of cores, but also its fragmentation to stellar masses also strongly depends on the ambipolar diffusion, the effectiveness of which remains a subject of discussion so far (Mouschovias \& Tassis, 2009). 

\sectionb{4}{CONCLUSION} 
We conclude that the magnetic field may delay and strongly retard the return of hydrogen of MCs into atomic phase, although the reliable theoretical or model estimates of this delay are absent. 

\thanks{This work was supported by Russian Foundation for Basic Research (project no. 12-02-00685-a) and FPS contract the Ministry of Education and Science 14.B37.21.0251}
\References
\refb Ballesteros-Paredes J. \& Hartmann, L., 2007, RMxAA, 43, 123
\refb Bash \& Dapp, ApJ, 716, 427, 2010
\refb Bertram E. et al., 2012, MNRAS, 420, 3163
\refb Blitz L. and Rosolowsky E., 2006, Astrophys. J., 650, 933
\refb Boquien M., Duc P.-A., Galliano F. et al. 2010, AJ, 140, 2124
\refb Boquien M., Duc P.-A., Wu Y. et al., 2009, Astron. J., 137, 4561
\refb Corbelli E., Bianchi S., Cortese L., 2012, A\&A, 542, 32
\refb Das M., Boone F., Viallefond, F.,2010, A\&A, 523, 63
\refb de Mello D., Torres-Flores S., Mendes de Oliveira C., 2008, Astron. J., 135, 319
\refb Dobbs, C. L.; Burkert, A.; Pringle, J. E., 2011, MNRAS, 413, 2935
\refb Egusa F., Koda J. and Scoville N., 2011, Astrophys. J., 726, 85
\refb Feigelson E.D., 1996, Astrophys. J., 468, 306
\refb Gratier P., 2012, A\&A, 542, 108
\refb Heitsch F., Mac Low M.-M., Klessen R. S., 2001, ApJ, 547, 280
\refb Heyer M., Krawczyk C., Duval J., Jackson J. M., 2009, ApJ, 699, 1092
\refb Hopkins P.; Quataert E.; Murray E, 2012, MNRAS, 421, p. 3488
\refb Jijina J., Myers P. C., Adams F. C., 1999, Astrophys. J. Suppl. Ser., 125, 161
\refb Karachentsev I. et al., 2011, MNRAS, 415, 31 
\refb Kasparova. A., 2012, Astron. Lett., 38, 63
\refb Koda J. et al., 2009, Astrophys. J., 700, 132
\refb Krumholz  M. R., McKee C. F., and Tumlinson J., 2006, Astrophys. J. 693, 216
\refb Kudoh T., Basu S., 2011, ApJ, 728, 123
\refb Li H.-B. et al., 2009, ApJ, 704, 891
\refb Li Hua-Bai, Henning, T., 2011, Nature, 479, 499
\refb Mouschovias T.Ch., Tassis T., 2009, MNRAS, 400, 15
\refb Mullan B., Konstantopoulos I., Kepley, A. et al., 2011, Astrophys. J., 731, 93
\refb Murray N., Quataert E., Thompson T. A., 2010, Astrophys. J., 709, 191
\refb Neff S. G., Thilker D. A., Seibert, M. et al., 2005, Astrophys. J., 619, 91
\refb Pfenniger D., Combes F., 1994, A\&A, 285, 94
\refb Surdin, 1981, Astron. Tsircular, 1178, 6
\refb Vazquez-Semadeni E., Banerjee R., Gomez G. et al., 2011, MNRAS, 414, 2511
\refb Vollmer B., Wong O., Braine J. et al., 2012, A\&A, 543A, 33
\refb Williams P. \& McKee F., 1997, Astrophys. J., 476, 166

\end{document}